\documentclass[%
reprint,
superscriptaddress,
 amsmath,amssymb,
 aps,
]{revtex4-1}

\usepackage{graphicx}
\usepackage{dcolumn}
\usepackage{bm}

\usepackage[]{units}
\usepackage{upgreek}
\usepackage{mathrsfs}

\usepackage{natbib}
\usepackage[colorlinks=true,citecolor=blue]{hyperref}
\begin{document}


\title{Energy partitioning and electron momentum distributions in intense laser-solid interactions}

\author{Joel Magnusson}
\email{joel.magnusson@chalmers.se}
\affiliation{Department of Physics, Chalmers University of Technology, SE-412 96 G\"oteborg, Sweden}

\author{Arkady Gonoskov}%
\affiliation{Department of Physics, Chalmers University of Technology, SE-412 96 G\"oteborg, Sweden}
\affiliation{Institute of Applied Physics, Russian Academy of Sciences, Nizhny Novgorod 603950, Russia}
\affiliation{Lobachevsky State University of Nizhni Novgorod, Nizhny Novgorod 603950, Russia}

\author{Mattias Marklund}
\affiliation{Department of Physics, Chalmers University of Technology, SE-412 96 G\"oteborg, Sweden}

\date{\today}

\begin{abstract}
Producing inward orientated streams of energetic electrons by intense laser pulses acting on solid targets is the most robust and accessible way of transferring the laser energy to particles, which underlies numerous applications, ranging from TNSA to laboratory astrophysics. Structures with the scale of the laser wavelength can significantly enhance energy absorption, which has been in the center of attention in recent studies. In this article, we demonstrate and assess the effect of the structures for widening the angular distribution of generated energetic electrons. We analyse the results of PIC simulations and reveal several aspects that can be important for the related applications.

\end{abstract}

\pacs{Valid PACS appear here}
\maketitle


\section{Introduction}
\label{sec:intro}

The interaction between high-intensity laser pulses and solids has been in the focus of researchers for a long period of time. Apart from being an experimental testbed for basic plasma phenomena, these interactions have several important applications, such as high harmonic generation (HHG) \cite{Teubner} or ion acceleration \cite{Daido,Macchi,Bulanov}. The latter has been widely studied due to the vast range of possible applications, e.g., biological or medical utilisations, of a table-top high-energy ion source. The overwhelming part of such studies relies on the target normal sheath acceleration (TNSA) scheme, in which the target's electrons are heated by the laser and then transported from the target's back side, creating an electrostatic acceleration field for the ions \cite{Roth,Mora,Cowan,Passoni}. 
There are many other schemes designed for ion acceleration (albeit exploiting different mechanisms), such as Coulomb explosion of clusters \cite{Ditmire,Kovalev1,Kovalev2}, double-layered targets \cite{Esirkepov1,Esirkepov2,Bulanov2}, collisionless schock acceleration \cite{Silva,Haberberger}, hole boring \cite{Schlegel}, light sail acceleration (or laser piston acceleration) \cite{Esirkepov3,Bulanov3,Henig,Kar}, and chirped standing wave acceleration \cite{Mackenroth}. 

The general simplicity of the TNSA mechanism and its robustness has made it easily accessible for experiments which by extension explains its popularity. Considerable effort has been put into the study of TNSA in order to fully understand and improve upon the basic scheme. To this end, various studies have been performed of specially designed targets and laser pulse shapes \cite{Flippo,Buffechoux,Burza,Gaillard,Markey,Pfotenhauer}.

However, it is well known that the TNSA scheme has several shortcomings, such as angular spread of the ions and the electrons being much faster than the ions. Many of these shortcomings can be ``engineered" away, i.e, by using a modified setup we can remedy, e.g., the angular spread. There are however basic restrictions that cannot be removed that easily. Most notably, the energy source is the laser pulse, and there is of course a limit on how much energy one can transfer from the laser to the target (and, therefore, to the ions). Theoretical and experimental studies show that the energy absorption can be significantly increased by structures on the surface \cite{Nodera,Cao,Zhao,Klimo,Margarone}, and the absorption can potentially be close to 100\% \cite{Blanco}. As a natural continuation of these studies, we consider here how the structures affect the partitioning of the absorbed energy between the low and high energy electrons as well as between their normal and transverse motion. Apart from enhancing TNSA, the obtained results can be useful for developing alternative ion acceleration schemes that utilize transverse streams of electrons \cite{Burza}, as well as for other applications such as creating streams of electrons for laboratory astrophysics experiments.

\section{Methods}
\label{sec:method}

In most attempts to improve our understanding of the TNSA scheme, the total absorbed laser energy and the resulting ion (typically protons) spectrum are considered. Analysing the total absorbed energy provides information about the initial stage of the process, of the interaction between the laser pulse and the plasma surface, while the ion spectrum essentially provides us with information integrated over the entire process. As important as these two sources of information are, this however does not provide us with the full picture. A large portion of the energy absorbed by the plasma at the front surface of the target will be `lost' instead of being transferred to the ions of interest: some hot electrons are transported away in the transverse directions, with some not even reaching the rear surface; electrons are backreflected and only transfer a small amount of their energy; and the transverse momentum of the hot electrons will not contribute to acceleration of the ions. 

While the diagnostic tools available for experiments remain limited, thus making experimental studies of the intermediary stages of the TNSA process difficult, modern computational codes with the ability to act as new diagnostic tools can be used in order to bridge the gap and provide new insights into this intermediate regime. Here we will make use of the particle-in-cell (PIC) code \textsc{Picador} \cite{Bastrakov}. 

In order to study the effect of microstructured targets on the unwanted transverse transport of hot electrons and to be able to put this in relation to, for example, the absorbed energy, we need a measure of the energy related to motion in the forward (normal) and transverse directions. We define these quantities given the set of criteria that they should (1) be proportional to the kinetic energy of the electrons, (2) reflect upon the direction of the electrons, (3) and be additive. The final criterion is important as we want to ensure that addition of the transverse and forward energies yields the total kinetic energy of the electrons accounted for. Thus, we will use the definitions

\begin{equation}\label{eq:Ex}
\mathscr{E}_x = \sum_{\mathrm{electrons}}\frac{p_x^2}{p_x^2+p_y^2} m_ec^2(\gamma-1),
\end{equation}
\begin{equation}\label{eq:Ey}
\mathscr{E}_y = \sum_{\mathrm{electrons}}\frac{p_y^2}{p_x^2+p_y^2}m_ec^2(\gamma-1),
\end{equation}
\begin{equation}\label{eq:E}
\mathscr{E} = \mathscr{E}_x + \mathscr{E}_y = \sum_{\mathrm{electrons}} m_ec^2(\gamma-1) = E_{\mathrm{kinetic}},
\end{equation}

where $m_e$ is the electron mass, $c$ the speed of light, $\gamma$ the Lorentz factor and $p_x$ and $p_y$ are the electron momenta in the $x$- and $y$-direction, respectively. Furthermore, as we are only interested in the energy of the electrons travelling in the forward direction, we only account for electrons transiting the virtual surface in the positive $x$-direction ($p_x > 0$).

\subsection{Targets}
\label{sec:targets}
\begin{figure}
\centering
\resizebox{\columnwidth}{!}{%
  \includegraphics{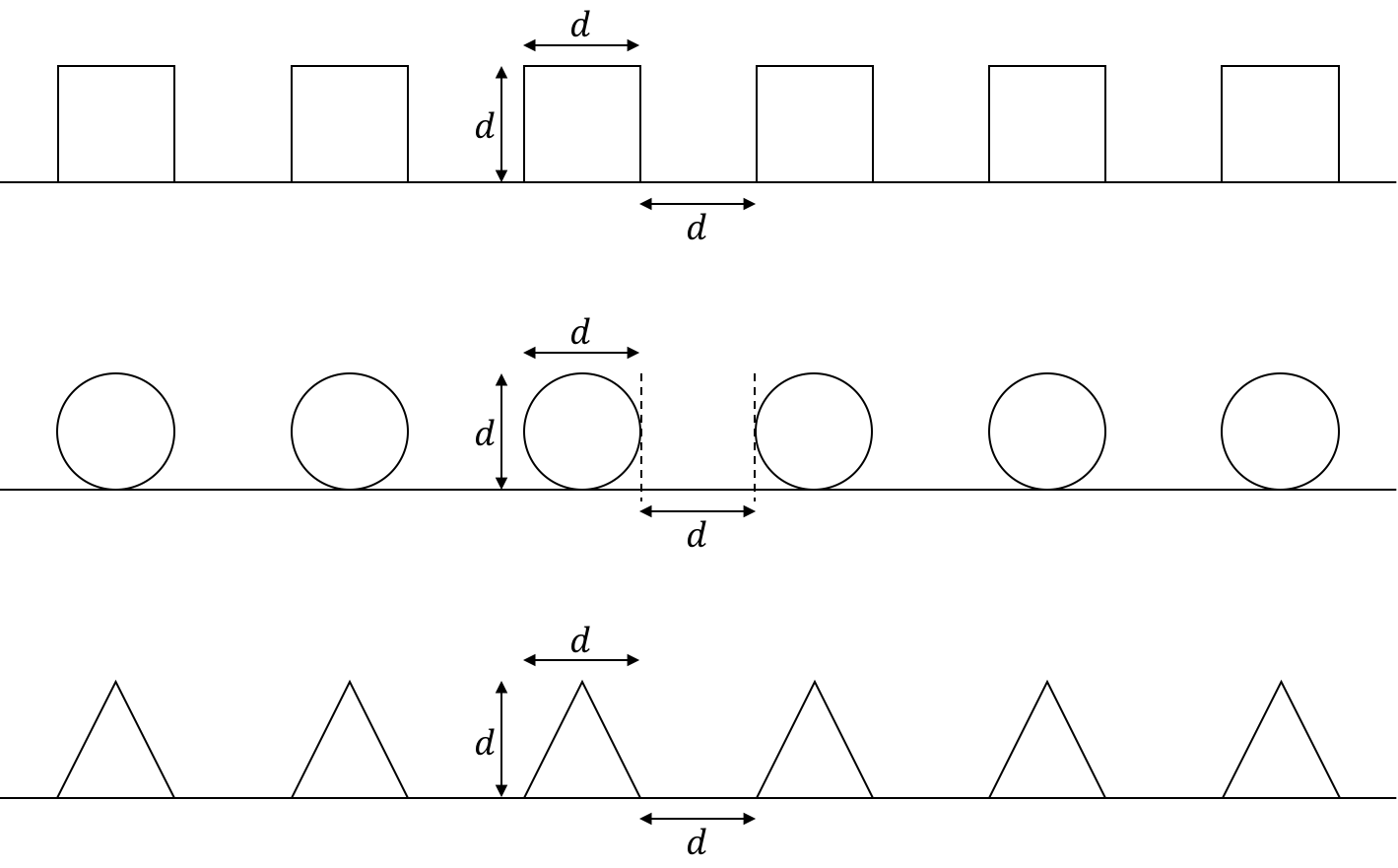}
}
\caption{The 2D geometry of the microstructured targets considered are completely described by a single scaling parameter, $d$, defining the maximum width and height as well as the minimal distance between the structures.}
\label{fig:microstructures}
\end{figure}

We study three different periodic microstructured designs; squares, triangles and circles (see Fig.\ \ref{fig:microstructures}). Each structure consists of a base shape of size $d \times d$, where we define $d$ as the linear scale of the structure. This structure is then periodically repeated across an otherwise flat surface, with a periodicity of $2d$. While the density and composition of such structures are in general independent of the rest of the target, they will for the purpose of this paper be identical to the rest of the target. A flat target will also be studied as a reference.

When studying structures of varying sizes we will limit ourselves to sizes in the range [$1/8\lambda$, $4\lambda$] in order to ensure that (1) the structures are sufficiently resolved by the space step ($dx \ll d$) and (2) the effects due to alignment remains negligible ($d < w_0$). Moreover the density profile of the targets studied will be sharp, meaning a negligible preplasma. This allows us to study a cleaner setup and to clearly find cause-and-effects in the setup(s). Adding a preplasma to these microstructured targets would certainly be of interest, but is outside the scope of the current study.

\begin{figure}
\centering
\resizebox{\columnwidth}{!}{%
  \includegraphics{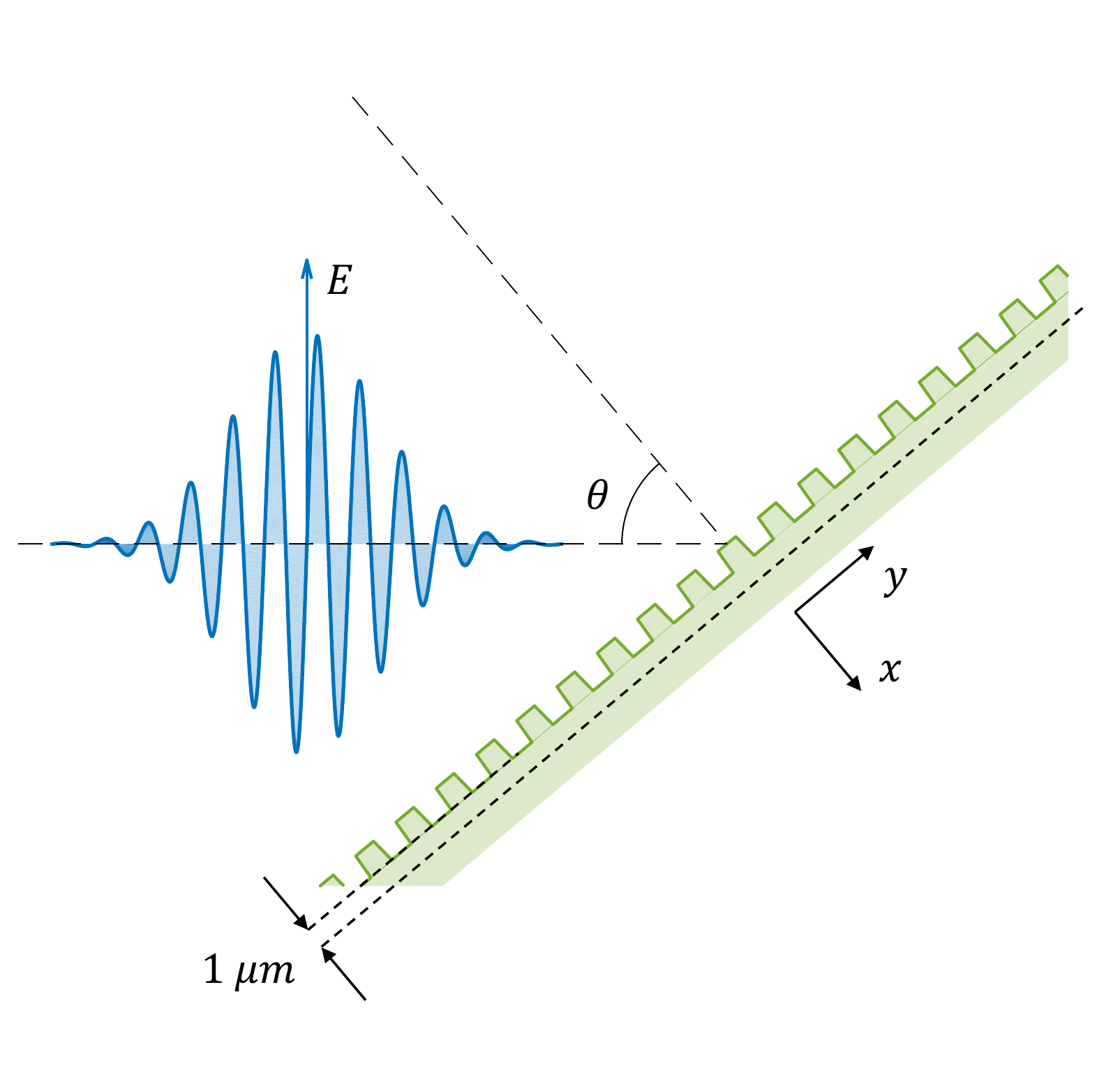}
}
\caption{The setup consists of a p-polarised gaussian laser pulse incident on a microstructured, semi-infinite and overdense plasma at an angle to the target normal of $\theta$. Placed $\unit[1]{\upmu m}$ inside the plasma, not counting the height of the microstructures, is a virtual surface at which distributions of transiting particles are calculated.}
\label{fig:setup}
\end{figure}

\begin{figure*}[t!]
\centering
\resizebox{0.75\textwidth}{!}{%
  \includegraphics{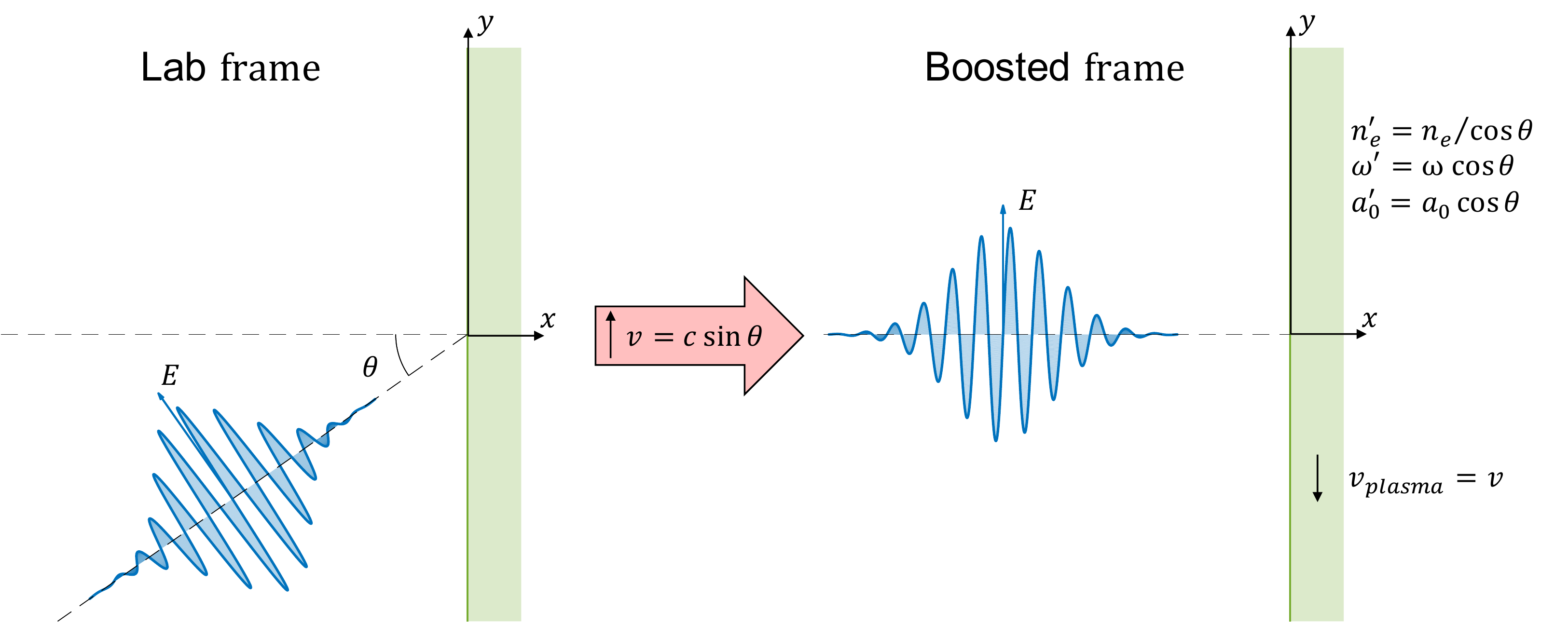}
}
\caption{A p-polarised laser pulse is incident on a flat plasma slab at an angle $\theta$. A Lorentz boost of $v=c\sin\theta$ in the $y$-direction, along the surface in the plane of incidence, results in a boosted frame where the laser pulse is instead incident normally on a slab of plasma streaming in the negative $y$-direction with velocity $v$. The electron plasma density, pulse carrier wavelength and pulse amplitude will be transformed according to the shown relations, where the primed quantities are that of the boosted frame and where we have retained the units of the lab frame.}
\label{fig:boostedframe}
\end{figure*}

\subsection{Laser pulse}
\label{sec:laser}
As we aim to use relevant laser parameters, available at most high-power laser facilities, we here consider a laser pulse of wavelength $\lambda = \unit[810]{nm}$ and energy $E = \unit[1]{J}$, with a Gaussian profile focused to a FWHM beam waist radius $w_0 = \unit[5]{\upmu m}$. Its peak amplitude is consequently given by $a_0=6.3$ with a corresponding peak intensity of $I=\unit[8.4 \cdot 10^{19}]{W/cm^2}$. The laser pulse is p-polarized in order to maximize the electron heating. 

The targets are modeled to be of solid density with a number density of $n_0 =  30n_{cr}$, with critical density $n_{cr} = m_e\omega_0^2/4\pi e^2$ and where $\omega_0$ is the laser pulse carrier frequency and $e$ is the electron charge.

\subsection{Numerical setup}
\label{sec:setup}
In order to study the aforementioned target designs we have performed two-dimensional simulations using the PIC code \textsc{Picador} \cite{Bastrakov}. We resort to two dimensional simulations in order to keep the computational cost at a feasible level, however, this restriction is not likely to significantly affect our result as
(1) the individual target geometries are two dimensional,
(2) the interaction between the pulse and the plasma and the subsequent motion of the electrons are sufficiently described in the plane of incidence, and
(3) two dimensions are sufficient for allowing important instabilities to form.

The targets are modeled as singly ionized plasmas consisting of electrons and heavy ions with mass $10 m_p$ and charge $-1e$, where $m_p$ is the proton mass.


\section{\label{sec:res}Results}
\label{sec:results}

When a flat foil is irradiated by a laser pulse at an angle $\theta$, the strong localized heating of the plasma will generate highly energetic electrons. The momentum distribution of these hot electrons can be calculated for highly idealized setups. As the physics remain unchanged under Lorentz transformations we may instead consider the completely equivalent system of a laser pulse irradiating a flat foil of streaming plasma at normal incidence, which can then be treated as one-dimensional (see Fig.\ \ref{fig:boostedframe}). 

In the boosted frame, $S^\prime$, the momenta of the electrons are given by

\begin{equation}
p_y^\prime =  -\gamma v/c = -\tan\theta,
\end{equation}
where the $\gamma$ is the Lorentz factor and the momenta are given in units of $m_e c$.

Moreover, by consideration of conservation of generalized momentum we have that
\begin{equation}
p_y^\prime + A_y^\prime =  \mathrm{const} = -\tan\theta, ~ p_x^\prime =  \mathrm{const} = P,
\end{equation}
where $A_y^\prime$ is the vector potential as seen in the boosted frame, written in units where the absolute value of the electron charge is 1. 

By applying a boost $v$ in the negative $y$-direction, we obtain the following expressions for the electron momenta in the lab frame
\begin{equation}
p_y = \frac{\sin\theta}{\cos^2\theta} \Big( \sqrt{1 +P^2\cos^2\theta} - 1 \Big), ~ p_x = P.
\end{equation}

\begin{figure}[t]
\centering
\resizebox{\columnwidth}{!}{%
  \includegraphics{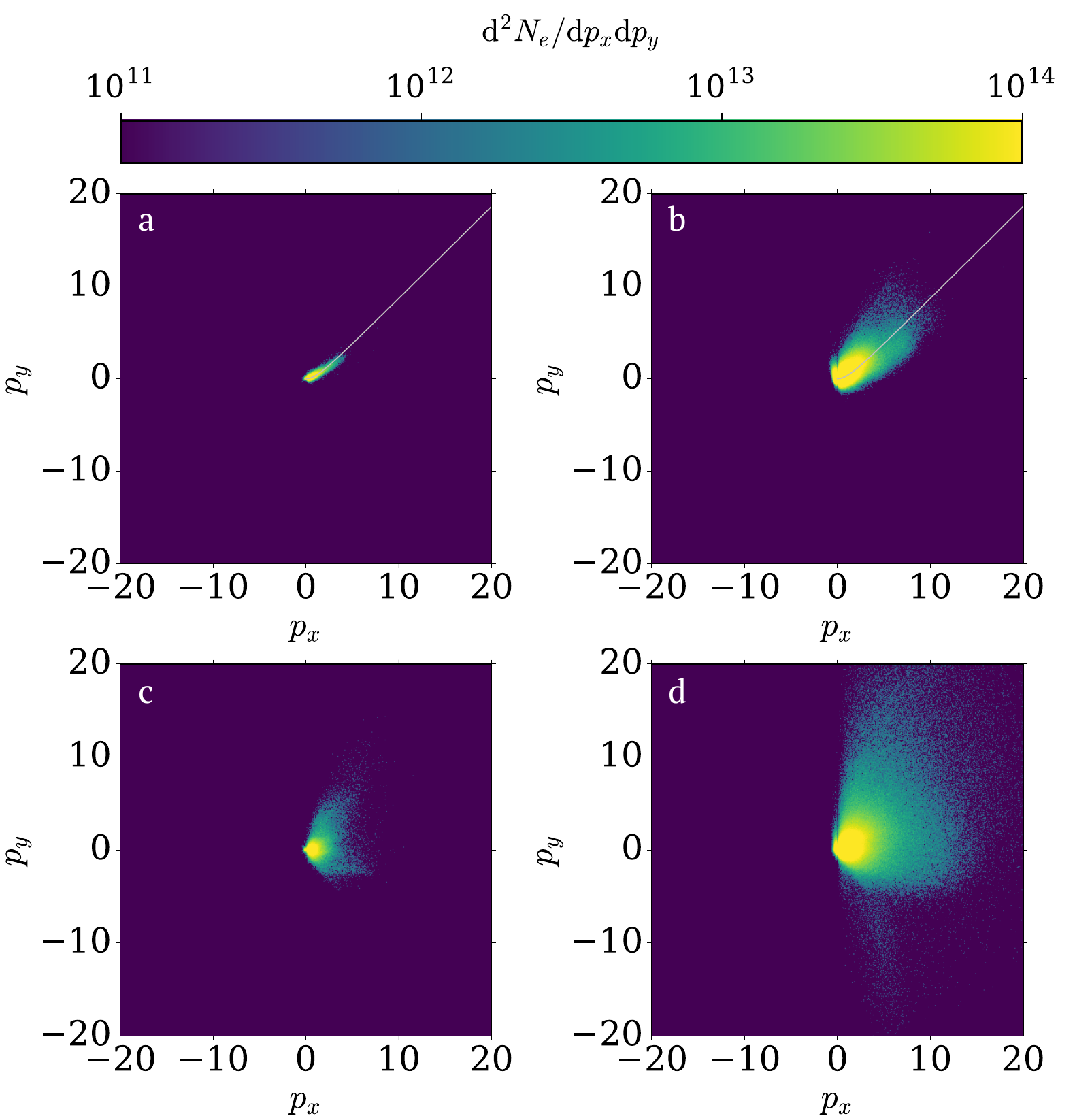}
}
\caption{The cumulative momentum space distribution of electrons transiting the virtual surface placed $\unit[1]{\upmu m}$ inside the plasma for a flat foil (a, b) and a foil with $d = \unit[0.5]{\upmu m}$ square microstructures (c, d) when irradiated by a laser pulse incident at $45^\circ$. The momentum relation predicted by conservation of generalized momenta for an idealized flat foil is indicated with a light-gray line (a, b). The distribution is shown at $t = \unit[225]{fs}$ (a, c) and $t = \unit[500]{fs}$ (b, d).}
\label{fig:pxpy}
\end{figure}

In two and three spatial dimensions, instabilities forming at the interaction surface will distort this and the collimation of the hot electrons will subsequently be decreased compared to the one dimensional case. This behaviour is easily recognizable in Fig. \ref{fig:pxpy} (a, b) where the momentum space distribution of the generated hot electrons closely follows the predicted shape at the earlier stages of the interactions. The distribution follows this general trend all throughout the interaction, however when integrating over the entire process it can be clearly seen that it gets spread out, as expected.

The introduction of microstructures to the surface can drastically change this behaviour. The introduction of the structures increases the number of incidence angles experienced by the pulse as it interacts with the surface and further breaks the homogeneity present in the transverse direction of flat targets. As a result the hot electrons will be generated with a much broader momentum space distribution, which can be seen from Fig. \ref{fig:pxpy} (c, d). Apart from increasing the absorption of laser energy by the plasma, the motion of the generated hot electrons will on average be more directed in the forward direction, thus decreasing the relative magnitude of the energy lost because of transverse transport of the hot electrons.

\begin{figure}[t]
\centering
\resizebox{\columnwidth}{!}{%
  \includegraphics{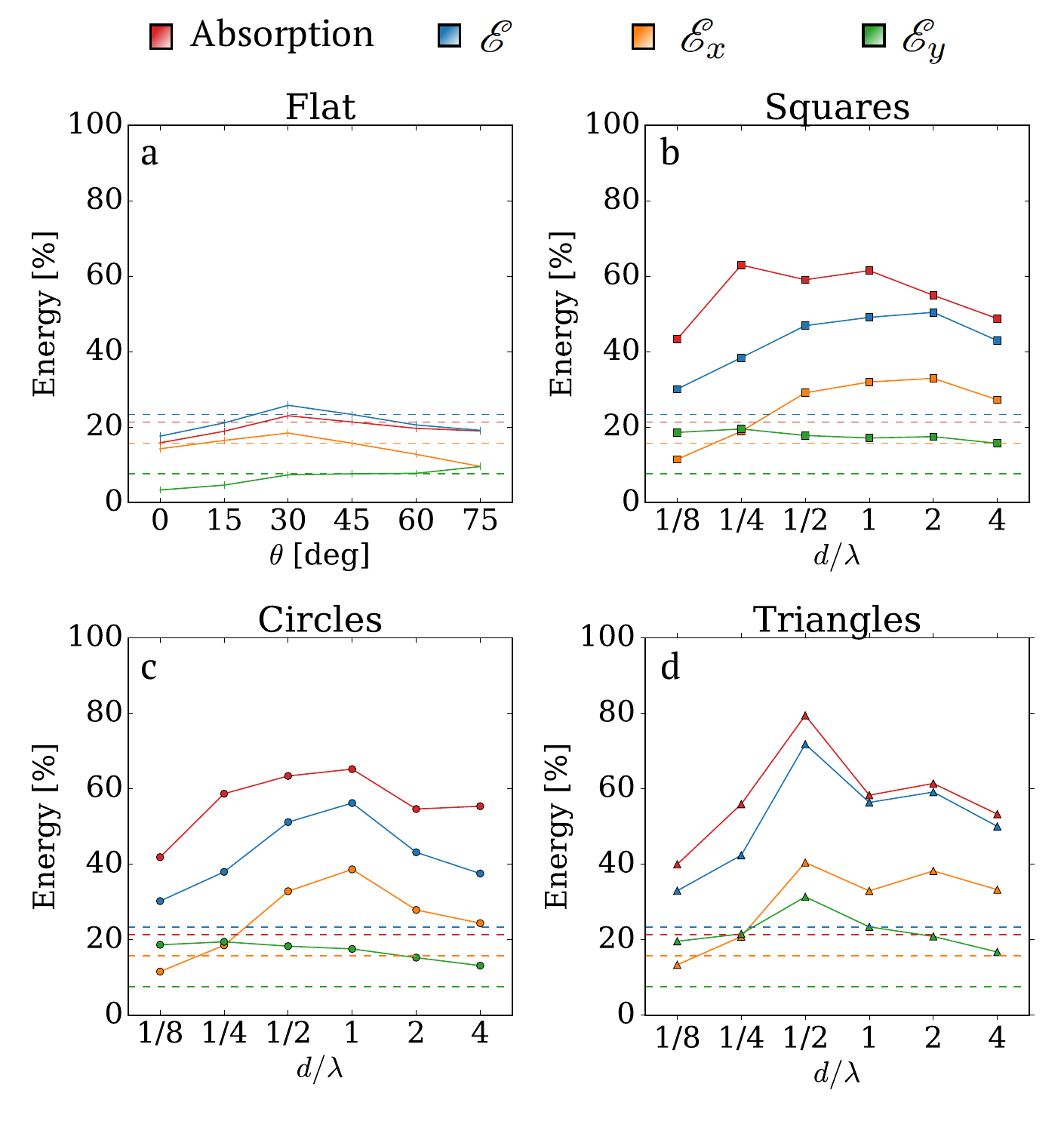}
}
\caption{The absorbed energy and the energy stored in forward and transverse motion of electrons when irradiating a flat foil at different angles of incidence (a) and microstructured targets irradiated at $45^\circ$ (b-d) are presented as percentages of the total laser energy. The result of a laser pulse incident at $45^\circ$ on a flat foil is indicated with dashed lines.}
\label{fig:results}
\end{figure}

The effect of varying the incidence angle on the energy of forward and transverse motion of hot electrons, as defined by Eq. \ref{eq:Ex} \& \ref{eq:Ey}, is presented in Fig. \ref{fig:results} (a). As expected the, the transverse motion energy ($\mathscr{E}_y$) steadily increases with the angle of incidence. The forward motion energy ($\mathscr{E}_x$) on the other hand first increases with the angle of incidence, peaks at $30^\circ$ and then decreases. This is mainly due to the improved coupling between the electric field of the laser and the plasma across the surface for oblique incidence. Furthermore, it should be pointed out that the total energy of electrons transiting the virtual surface is slightly larger than the total absorbed laser energy, as seen in the figure, due to double counting of refluxing particles mainly associated with the local heating of the plasma by the main pulse.

Instead, looking at how these quantities are affected by microstructures of varying shape and size, we see that the transverse motion energy remains relatively unchanged as the linear scale of the structures increase. An interesting exception to this is the triangular structures, which instead displays a strong peak at a size of $d = 1/2\lambda$. What is more important however, is the fact that the forward energy is greatly improved for $1/2 \lesssim d / \lambda \lesssim 2$ for all three cases. Thus, the additional energy transfered to the hot electrons is mainly contributing to their forward motion.

\begin{figure}
\centering
\resizebox{\columnwidth}{!}{%
  \includegraphics{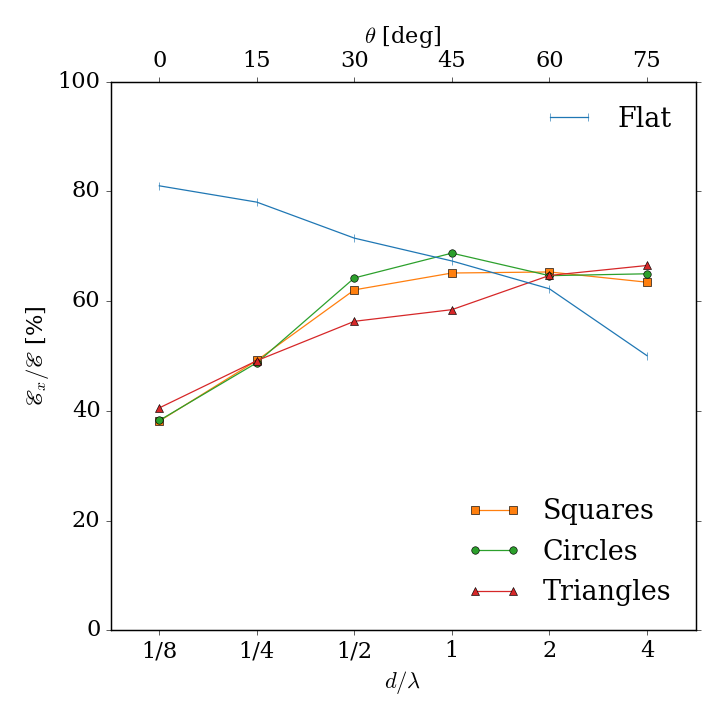}
}
\caption{The energy stored in forward motion of electrons transiting a virtual surface located $\unit[1]{\upmu m}$ inside the plasma slab presented as percentage of the total kinetic energy of the transiting electrons.}
\label{fig:Ex}
\end{figure}

Since the relative energy of forward motion can be seen to increase with the addition of the microstructures, it is interesting to also compare the targets using this measure, as it provides information about the relative amount of energy of the electrons expected to be 'useful' to applications such as TNSA, where useful is taken to mean: transferable to the ions at the rear surface. As can be seen from Fig. \ref{fig:Ex} the relative energy of forward motion for the three structured designs follows a very similar trend, despite their differences in absolute energy, as seen in Fig. \ref{fig:Ex} (b-d). Furthermore the relative energy of forward motion displays a decreasing trend with increasing angle of incidence and with $0^\circ$ (being the optimal angle), as expected.

Moreover, a clear discrepancy between the total absorption and total energy of the transiting hot electrons can be seen for mainly sub-wavelength square and circular structures, presented in Fig. \ref{fig:results} (b, c). Taking into consideration that at least some double counting occurs in calculating the transit energy makes this difference all the more significant. This clearly shows that a significant portion of the absorbed energy does not get carried across the virtual surface, by the electrons. A fraction of this energy can be found in the semi-static fields forming at the plasma interface, but the lion's part can be found as kinetic energy of hot electrons trapped by a shock front, thus being prevented from travelling further into the plasma. It is also interesting to note that this behaviour is much less pronounced with the triangular structures which opens up for the possibility of controlling it by smart design choices.

\section{Conclusions}
\label{sec:outlook}
In this paper we have made a detailed parametric study of the TNSA scheme, under the influence of microstructures on the front side of the target. First, we analyses the limitations of the scheme in terms of laser energy absorption and spectral properties of the generated hot electrons. We found, expectedly, that the absorption indeed was affected by the microstructures, and that the size of the structures affected the absorption. However, we also found that the hot electron distribution was significantly affected by the target structures. 

We would like to stress the importance of the latter findings. As stated, it is well-known that front surface structures affects the general absorption properties of the target. But not only can the structures increase the total absorption, they can also decrease the transverse transport of the hot electrons, thus directing a larger fraction of the electrons in the forward direction. This can enable a stronger acceleration of the ions of interest. 

Our findings also points us to the following interesting conclusions:
\begin{itemize}
    \item there is a limited view of the TNSA scheme when \emph{only} considering increased absorption: it is only possible to obtain $\sim$ two times stronger absorption by structuring the targets, as given in this paper, but for most applications we would require the energy transfer to go much further;
    \item the partitioning of energy in the hot electron distribution is important; only the electron momenta in the $x$-direction can be considered beneficial for ion acceleration, as large transverse momenta will represent an energy loss channel in this sense;
    \item however, a large angular divergence of electrons leads to small ion angular divergence, but also a small cutoff ion energy, and a smaller electron angular divergence leads to larger ion divergence, but also increased cutoff energy; thus, there is a trade-off. 
\end{itemize}

As final conclusion, we find that it is of central importance to control the electron distribution. For this purpose, improvements can be made in the stages after the electron heating, using, e.g., strong guiding magnetic fields or mass limited targets. There is also the possibility of using targets cleverly designed to take advantage of the directionality of the hot electrons or to guide their transverse motion in order to increase the energy transfered from the electrons to the ions (see e.g. \cite{burza.njp.2011}). There are thus ample opportunities for future studies and improvements of the TNSA scheme.

\begin{acknowledgments}
The authors would like to thank the \textsc{Picador} development team, and S. Bastrakov in particular, for their invaluable technical support and acknowledge the financial support by the Wallenberg Foundation through the grant ``Plasma based compact ion sources'' (PLIONA). The simulations were performed on resources provided by the Swedish National Infrastructure for Computing (SNIC) at HPC2N.

\end{acknowledgments}

\bibliography{references}


\hfill
\end{document}